\documentclass[conference,a4paper]{APSIPA2021}
\usepackage{cite}
\usepackage[numbers]{natbib}
\usepackage[colorlinks,citecolor=black]{hyperref}
\usepackage{multirow}
\usepackage{amsmath}
\usepackage[psamsfonts]{amssymb}
\usepackage{amsfonts}
 \usepackage{booktabs}
\usepackage{amsxtra}
\usepackage{threeparttable}
\usepackage{graphicx}
\usepackage{float}
\usepackage{bm}
\usepackage{geometry}
\usepackage[normalem]{ulem}
\useunder{\uline}{\ul}{}
\usepackage{makecell}

\begin{document}

\title{FlowGRPO for Vocal Separation: From Reward Analysis to Reward Design}

\author{
Lishi Zuo, Chong-Xin Gan, Zezhong Jin, Youzhi Tu, Man-Wai Mak, and KongAik Lee\\
\texttt{Dept. of Electrical and Electronic Engineering,}\\
\texttt{ Hong Kong Polytechnic University, Hong Kong SAR, China}\\
\texttt{\{lishi.zuo, chong-xin.gan, zezhong.jin\}@connect.polyu.hk}\\
\texttt{\{youztu, man.wai.mak, kong-aik.lee\}@polyu.edu.hk}
}

\maketitle
\thispagestyle{empty}
\begin{abstract}

The Flow Matching objective supervises the velocity field at intermediate states, whereas waveform metrics are evaluated after decoding and may be non-differentiable. 
To align Flow Matching models with waveform-level and perceptual quality objectives during post-training, FlowGRPO optimizes their trajectories using endpoint rewards.
In this work, we focus on music separation and use multi-resolution short-time Fourier transform (MRSTFT), scale-invariant signal-to-distortion ratio (SI-SDR), and Virtual Speech Quality Objective Listener (ViSQOL) as reward components. Before policy optimization, we analyze the relationship between data quality factors and the within-group gap between the maximum and mean rollout rewards for each component. This analysis motivates a revised ViSQOL reward that matches channel-wise root-mean-square (RMS) levels and evaluates the interval from the first to the last active frame, retaining inactive frames within that interval. We also apply a sample-quality filter based on predefined waveform checks. Together, the revised reward and quality filter enable FlowGRPO to improve the pretrained vocal separator in both separation performance and perceptual quality.
\end{abstract}

\section{Introduction}
Generative models are used for speech separation, speech enhancement, and target speaker extraction~\cite{diffusion_source_separation,diffusion_speech_enhancement,gen_target_speaker_extraction}. Recently, flow Matching has been applied to speech generation and audio source separation~\cite{fm_audio_1,fm_audio_2}.

Aligning a pretrained Flow Matching model to a waveform objective remains difficult. This is because the Flow Matching objective supervises the velocity field at intermediate flow states, whereas downstream metrics evaluate only the final waveform~\cite{lipman2023flow,flowgrpo}. Also, many speech and audio metrics are non-differentiable or depend on pretrained models and signal-processing pipelines~\cite{wer_training,valle,metricgan,sdr_bss_eval}. In latent Flow Matching, the variational autoencoder (VAE) decoder~\cite{vae} maps the generative trajectory to the waveform, so these metrics cannot be applied directly to latent states.

Reinforcement learning (RL) provides a way to optimize non-differentiable metrics by treating them as scalar rewards, without requiring their gradients~\cite{koizumi2017,koizumi2018,f5rtts}. Specifically, FlowGRPO applies Group Relative Policy Optimization (GRPO)~\cite{grpo} to Flow Matching trajectories and uses endpoint metrics as rewards~\cite{flowgrpo}. Since these rewards guide policy updates to the Flow Matching model, reward selection and design are critical to the effectiveness of post-training optimization. To avoid costly policy optimization experiments, we investigate the following question: \textbf{\textit{Can reward analysis before policy optimization inform reward design for FlowGRPO?}}

We investigate reward design for FlowGRPO-based vocal separation by analyzing the within-group best-to-mean gaps of three reward components: multi-resolution short-time Fourier transform (MRSTFT), scale-invariant signal-to-distortion ratio (SI-SDR)~\cite{sisdr}, and Virtual Speech Quality Objective Listener (ViSQOL)~\cite{visqol}. For each component, the gap is defined as the difference between the highest and mean rollout rewards for the same mixture. We find that the ViSQOL gap is negatively associated with the reference vocal's inter-channel root-mean-square (RMS) difference and edge-silence ratio. These findings motivate a revised ViSQOL reward that matches RMS levels per channel and evaluates the interval from the first to the last active frame while retaining internal inactive frames. We also apply a sample-quality filter based on predefined waveform checks. Experiments show that, together, the revised reward and quality filter enable FlowGRPO to improve the pretrained vocal separator in both separation performance and perceptual quality.

Our contributions are:
\begin{itemize}
\item \textbf{Reward analysis.} We select the multi-resolution short-time Fourier transform (MRSTFT), SI-SDR, and ViSQOL as vocal-separation rewards and measure their within-group best-to-mean gaps before policy optimization.
\item \textbf{Reward design.} From this analysis, we revise ViSQOL with channel-wise RMS matching and scoring from the first to the last active frame, and combine it with MRSTFT and SI-SDR. We also apply a sample-quality filter based on fixed waveform checks.
\item \textbf{Empirical results.} FlowGRPO with the revised reward and the filter improves the pretrained vocal separator on all five reported metrics (Table~\ref{tab:reward_ablation}).
\end{itemize}

\section{Background and Related Work}
\label{sec:bg}

\subsection{Flow Matching and Audio Source Separation}
\label{sec:bg:fm}

Diffusion models are widely used for speech and audio generation~\cite{diffwave,wavegrad,gradtts}, and have been extended to speech enhancement and source separation~\cite{diffusion_speech_enhancement,diffusion_source_separation}. Flow Matching instead learns a continuous-time velocity field that transports samples from a prior to the data distribution~\cite{lipman2023flow}. It has been applied to speech generation~\cite{fm_audio_1}, audio source separation~\cite{fm_audio_2}, and joint vocal-accompaniment separation~\cite{zuo2026decoupled}.

Vocal separation aims to extract the vocal stem from a music mixture. Conventional discriminative approaches estimate the vocal stem directly from the mixture, operating either on waveforms~\cite{demucs} or on frequency subbands using recurrent and Transformer architectures~\cite{bsrnn,bsroformer}. More recently, generative approaches have formulated vocal separation as a diffusion process conditioned on the mixture~\cite{diff_singing_vocal}. Rather than introducing a new separator architecture, we study reward-based post-training of a pretrained latent Flow Matching separator.

\subsection{Reinforcement Learning and Reward Design for Flow Matching}
\label{sec:bg:rl}

Reinforcement learning enables the optimization of non-differentiable speech and audio metrics by treating them as scalar rewards~\cite{koizumi2017,koizumi2018,f5rtts}. Building on this idea, FlowGRPO applies GRPO to Flow Matching models, optimizing continuous generative trajectories using task-specific endpoint rewards~\cite{flowgrpo}. Subsequent work has explored more targeted stochastic exploration strategies. Fast-GRPO confines stochastic updates to a short time window~\cite{prismaudio}, while TempFlow-GRPO shows that exploration at early steps has a substantially greater impact on terminal rewards than exploration at later steps~\cite{tempflowgrpo}. Similarly, FlowTTS-GRPO restricts stochastic differential equation (SDE) sampling to early steps~\cite{flowttsgrpo}.

Because rewards serve as proxies for task objectives, differences in their scales, sensitivities, and correlations can affect multi-reward optimization, while reward misspecification may lead to reward hacking~\cite{reward_hacking_1,reward_hacking_2}. We therefore analyze within-group reward variation before policy optimization and use the observed associations to inform reward redesign.

\section{Methodology}
\label{sec:method}

We formulate latent vocal separation and the FlowGRPO post-training objective. We then describe the reward analysis used before training and the reward design that follows from it.\looseness-1

\subsection{Vocal Separation in the VAE Latent Space}
\label{sec:latent-sep}

We perform vocal separation in the latent space of a frozen, pretrained VAE, using its encoder $\mathrm{Enc}$ and decoder $\mathrm{Dec}$. Since separation metrics are defined on waveforms, we compute rewards by decoding the terminal latent representation. Let $\bm{x}_{\mathrm{mix}}$ denote the mixture waveform and $\bm{x}$ the reference vocal waveform. Both are encoded into latent representations with $C$ channels and $L$ frames:
\begin{equation}
\bm{Z}_{\mathrm{mix}}
=
\mathrm{Enc}(\bm{x}_{\mathrm{mix}}),
\qquad
\bm{Z}_{1}
=
\mathrm{Enc}(\bm{x})
\in\mathbb{R}^{C\times L}.
\label{eq:vae-enc}
\end{equation}
Here, $\bm{Z}_{1}$ denotes the target vocal latent at time $t=1$. The symbols $\bm{0}$ and $\mathbf{I}$ denote the zero element and identity covariance matrix, respectively, in $\mathbb{R}^{C\times L}$. Following the standard flow matching formulation~\cite{lipman2023flow}, Gaussian noise $\bm{Z}_{0} \sim \mathcal{N}(\bm{0},\mathbf{I})$ is transported to $\bm{Z}_{1}$ along a linear path.
\begin{equation}
\bm{Z}_{t}
=
(1-t)\bm{Z}_{0}
+
t\bm{Z}_{1},
\qquad
t\in[0,1],
\label{eq:fm-path}
\end{equation}
with a target velocity $\bm{v}=\bm{Z}_{1}-\bm{Z}_{0}$. A conditional flow matching model parameterized by $\theta$ predicts the velocity $\bm{v}_{\theta}(\bm{Z}_{t},t,\bm{Z}_{\mathrm{mix}})$ and is pretrained to estimate the velocity $\bm{v}$. At inference time, we integrate the predicted velocity field from $\bm{Z}_{0}$ to obtain the estimated endpoint $\hat{\bm{Z}}_{1}$, from which the vocal waveform is reconstructed as
\begin{equation}
\hat{\bm{x}}
=
\mathrm{Dec}(\hat{\bm{Z}}_{1}).
\label{eq:decode}
\end{equation}
The waveform-level reward components are computed from $\hat{\bm{x}}$ and the reference vocal $\bm{x}$.

\subsection{FlowGRPO}
\label{sec:flowgrpo}

We post-train $\theta$ using FlowGRPO~\cite{flowgrpo}, while keeping $\mathrm{Enc}$ and $\mathrm{Dec}$ frozen. Rollouts are generated using the old velocity field $\bm{v}_{\theta_{\mathrm{old}}}$, which remains fixed throughout each policy update. The probability-flow ordinary differential equation (ODE)
\begin{equation}
\mathrm{d}\bm{Z}_{t}
=
\bm{v}_{\theta}(\bm{Z}_{t},t,\bm{Z}_{\mathrm{mix}})\,\mathrm{d}t
\end{equation}
defines a deterministic transition from each state to its next state and therefore does not provide the non-degenerate transition probabilities required for policy optimization. To obtain tractable transition probabilities, FlowGRPO replaces the probability-flow ODE with a stochastic differential equation (SDE) that preserves the same marginal distributions at each time $t$~\cite{flowgrpo}. The SDE is discretized into $N$ steps of size $\Delta t=1/N$, with diffusion coefficient $\sigma_t=a\sqrt{(1-t)/t}$, where $a>0$ controls the noise scale.
The SDE transition follows a gaussian distribution and is given by
\begin{align}
\bm{Z}_{t+\Delta t}
&=
\boldsymbol{\mu}_{\theta,t}
+
\sigma_t\sqrt{\Delta t}\,\bm{\xi},
\qquad
\bm{\xi}\sim\mathcal{N}(\bm{0},\mathbf{I}),
\label{eq:sde-step}
\end{align}
where the transition mean is
\begin{align}
\boldsymbol{\mu}_{\theta,t}
&=
\bm{Z}_{t}
+
\bigg[
\bm{v}_{\theta}(\bm{Z}_{t},t,\bm{Z}_{\mathrm{mix}})
\nonumber\\
&\quad
+
\frac{\sigma_t^2}{2(1-t)}
\big(-\bm{Z}_{t}+t\,\bm{v}_{\theta}(\bm{Z}_{t},t,\bm{Z}_{\mathrm{mix}})
\big)
\bigg]\Delta t,
\label{eq:sde-mean}
\end{align}
The second term in the transition mean ensures that the SDE preserves the marginal distributions of the corresponding probability-flow ODE. 
Replacing $\bm{v}_{\theta}$ with the old velocity field $\bm{v}_{\theta_{\mathrm{old}}}$ gives the transition distribution $p_{\theta_{\mathrm{old}}}$ used to evaluate the rollout probabilities. 

Because the reward is evaluated at the decoded endpoint, applying the SDE at every step is unnecessary. As in prior FlowGRPO methods~\cite{flowgrpo,prismaudio}, noise is injected only inside a window \(\mathcal{W}\) of \(K\) consecutive steps, \(1\le K\le N-1\). The window starts at step $n_0$, drawn at random, and covers the times $\mathcal{W}=\{n_0\Delta t,\ldots,(n_0+K-1)\Delta t\}$. We integrate one trajectory with the ODE of $\theta_{\mathrm{old}}$ up to step $n_0$, then replicate its state into a group of $G$ trajectories. Independent noise is injected only inside \(\mathcal{W}\). After the window, every trajectory is integrated to $t=1$ with that ODE. Trajectories in a group therefore share the ODE path before \(\mathcal{W}\).

For mixture $g$ and trajectory $i\in\{1,\ldots,G\}$, we decode $\hat{\bm{Z}}_{1}^{(i)}$ with Eq.~\ref{eq:decode}. Let $r_{g,i}$ denote its scalar separation reward; Section~\ref{sec:reward} combines the reward components into this scalar. The group-relative advantage follows GRPO~\cite{grpo},
\begin{equation}
\hat{A}^{(i)}
=
\frac{
r_{g,i}-\bar r_g
}{
s_g+\delta_A
},
\label{eq:advantage}
\end{equation}
where $\bar r_g$ and $s_g$ are the within-group mean and standard deviation, and $\delta_A=10^{-4}$ prevents division by zero. Subtracting $\bar r_g$ removes the reward level shared inside the group, and dividing by $s_g$ keeps a high-variance group from dominating the update. Trajectory $i$ receives this advantage at every step $t\in\mathcal{W}$. The importance ratio is
\begin{equation}
\rho_t^{(i)}(\theta)
=
\frac{
p_\theta\big(\bm{Z}_{t+\Delta t}^{(i)}\mid\bm{Z}_{t}^{(i)},\bm{Z}_{\mathrm{mix}}\big)
}{
p_{\theta_{\mathrm{old}}}\big(\bm{Z}_{t+\Delta t}^{(i)}\mid\bm{Z}_{t}^{(i)},\bm{Z}_{\mathrm{mix}}\big)
}.
\label{eq:ratio}
\end{equation}
We set $\bar{\rho}_t^{(i)}(\theta)=\mathrm{clip}(\rho_t^{(i)}(\theta),1-\epsilon,1+\epsilon)$ for $\epsilon>0$, which limits the size of each policy update. We maximize the training objective
\begin{equation}
\begin{aligned}
\mathcal{J}(\theta)
&=
\mathbb{E}
\Bigg[
\frac{1}{G}\sum_{i=1}^{G}
\frac{1}{K}\sum_{t\in\mathcal{W}}
\min\big(\rho_t^{(i)}(\theta)\hat{A}^{(i)},\,
\bar{\rho}_t^{(i)}(\theta)\hat{A}^{(i)}\big)
\Bigg].
\end{aligned}
\label{eq:grpo}
\end{equation}

\subsection{Within-Group Reward Analysis}
\label{sec:reward-analysis}

GRPO updates the policy based on relative reward comparisons among rollouts generated from the same mixture. Therefore, the ability of a reward component to distinguish between rollouts within a group is important. To examine whether this ability depends on data quality factors, we analyze the within-group variation of three reward components: MRSTFT, SI-SDR, and ViSQOL.

For each mixture $g$, the old policy $\theta_{\mathrm{old}}$ generates $G$ rollouts. Let $R_{g,i}^{(m)}$ denote reward component $m$ for rollout $i$. All three reward components are defined such that higher values indicate better separation quality. We quantify this variation using the best-to-mean gap:
\begin{equation}
d_g^{(m)}
=
\max_{1\le i\le G} R_{g,i}^{(m)}
-
\frac{1}{G}\sum_{i=1}^{G}R_{g,i}^{(m)},
\label{eq:reward-gap}
\end{equation}
where a larger gap indicates a greater difference between the best rollout and the group average.

We then examine how this gap relates to five data quality factors measured over the valid target duration: RMS level, activity ratio above $-40$ decibels relative to full scale (dBFS), edge-silence ratio, stereo RMS gap in decibels (dB), and channel onset lag. Let $q$ index a factor and $q_g$ denote its value for mixture $g$. We use $z(\cdot)$ to denote standardization across mixtures. Let $\alpha_{q,m}$ denote the coefficient of factor $q$ for reward component $m$. We regress $z(d_g^{(m)})$ against $z(q_g)$, controlling for duration and the remaining level and stereo factors in $\mathbf{u}_{g,q}$. Activity ratio and edge-silence ratio are excluded from each other's control variables. The resulting coefficients are reported in Table~\ref{tab:reward_gap}.

\begin{table}[t]
\centering
\caption{Pre-training analysis of the within-group gap on $1500$ training mixtures. Entries are standardized regression coefficients $\alpha_{q,m}$. Bold denotes $|\alpha_{q,m}|>0.05$.}
\label{tab:reward_gap}
\begin{tabular}{lccc}
\toprule
Factor & $R_{\mathrm{MRSTFT}}$ & $R_{\mathrm{SI\text{-}SDR}}$ & $R_{\mathrm{ViSQOL}}$ \\
\midrule
RMS level          & $-0.005$ & $\mathbf{-0.062}$ & $\mathbf{-0.077}$ \\
Activity ratio     & $0.030$  & $0.006$  & $\mathbf{0.142}$ \\
Edge-silence ratio & $-0.018$ & $-0.015$ & $\mathbf{-0.063}$ \\
Stereo RMS gap     & $-0.048$ & $-0.012$ & $\mathbf{-0.107}$ \\
Channel onset lag  & $-0.016$ & $-0.029$ & $-0.008$ \\
\bottomrule
\end{tabular}
\end{table}

\begin{table*}[!t]
\centering
\renewcommand{\arraystretch}{1.08}
\caption{Vocal-separation quality. The reward components are combined using Eqs.~\ref{eq:reward-scalar} and~\ref{eq:advantage}. The ViSQOL column always reports the original metric.}
\label{tab:reward_ablation}
\setlength{\tabcolsep}{5.5pt}
\begin{tabular}{c|cl| c |ccccc}
\toprule
Row & Stage & Reward components & Filter & ViSQOL$\uparrow$ & SI-SDR$\uparrow$ & MRSTFT$\downarrow$ & SDR$\uparrow$ & PESQ$\uparrow$\\
\midrule
1 & Pretrain & -- & No & 3.525 & 5.927 & 1.148 & 6.946 & 2.076 \\
2 & FlowGRPO & $R_{\mathrm{MRSTFT}}$ & No & 3.594 & 6.858 & \textbf{1.067} & 7.991 & 2.103 \\
3 & FlowGRPO & $R_{\mathrm{MRSTFT}} + R_{\mathrm{SI\text{-}SDR}} + R_{\mathrm{ViSQOL}_{\mathrm{act}}}$ & No & 3.591 & 6.687 & 1.108 & 7.887 & 2.079\\
4 & FlowGRPO & $R_{\mathrm{MRSTFT}} + R_{\mathrm{SI\text{-}SDR}} + R_{\mathrm{ViSQOL}}$ & Yes & 3.641 & 6.900 & 1.098 & 8.244 & 2.066\\
5 & FlowGRPO & $R_{\mathrm{MRSTFT}} + R_{\mathrm{SI\text{-}SDR}} + R_{\mathrm{ViSQOL}_{\mathrm{act}}}$ & Yes & \textbf{3.675} & \textbf{7.045} & 1.084 & \textbf{8.293} & \textbf{2.112}\\
\bottomrule
\end{tabular}
\end{table*}

\begin{figure*}[t]
\centering
\includegraphics[width=\textwidth]{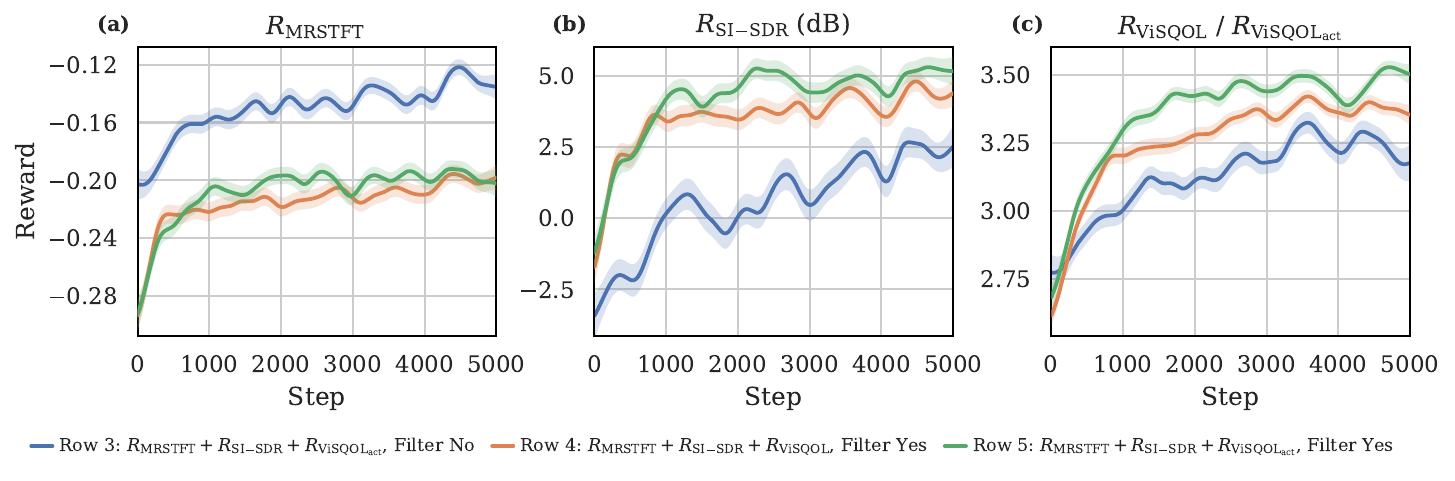}
\caption{Training rewards for Rows 3--5, plotted as the raw reward components before batch and group normalization. Panel (c) shows $R_{\mathrm{ViSQOL}}$ for Row 4 and $R_{\mathrm{ViSQOL}_{\mathrm{act}}}$ for Rows 3 and 5. Shading denotes $\pm1$ standard error.}
\label{fig:reward_curves}
\end{figure*}

A larger $|\alpha_{q,m}|$ indicates a stronger partial association between factor $q$ and the within-group gap of reward component $m$, after controlling for $\mathbf{u}_{g,q}$. For the bolded factors in Table~\ref{tab:reward_gap}, ViSQOL has a larger absolute coefficient than MRSTFT and SI-SDR. These results suggest that the within-group gap of ViSQOL is more strongly associated with certain data quality factors. Based on these observations, we revise the ViSQOL computation and introduce a sample-quality filter.

\subsection{Reward Design and Sample Filtering}
\label{sec:reward}

\textbf{Reward redesign.} We revise ViSQOL and leave MRSTFT and SI-SDR unchanged. The revision is applied to each channel separately. Let $\bm{x}_c$ and $\hat{\bm{x}}_c$ be the reference and generated waveforms of channel $c$. Let $\mathcal{F}$ be the interval from the first active frame of $\bm{x}_c$ to its last active frame. Inactive frames inside $\mathcal{F}$ are kept, so pauses within the vocal are still scored. We scale $\hat{\bm{x}}_c$ so that its RMS on $\mathcal{F}$ equals that of $\bm{x}_c$, $\tilde{\bm{x}}_c=\hat{\bm{x}}_c\frac{\mathrm{RMS}_{\mathcal{F}}(\bm{x}_c)}{\mathrm{RMS}_{\mathcal{F}}(\hat{\bm{x}}_c)}$.
$\mathrm{ViSQOL}_{\mathrm{act}}$ is the average over channels of the ViSQOL score between $\tilde{\bm{x}}_c$ and $\bm{x}_c$ on $\mathcal{F}$. SI-SDR already removes a global gain mismatch through its optimal projection, so this RMS scaling would not change its score.

The settings in Table~\ref{tab:reward_ablation} use subsets of MRSTFT, SI-SDR, and either the original or the revised ViSQOL. The proposed multi-reward combines MRSTFT, SI-SDR, and $\mathrm{ViSQOL}_{\mathrm{act}}$. Here, $R_{\mathrm{SI\text{-}SDR}}$ is the SI-SDR score, $R_{\mathrm{ViSQOL}_{\mathrm{act}}}$ is the revised ViSQOL reward, and $R_{\mathrm{ViSQOL}}$ is the original score. The subscript $\mathrm{act}$ denotes scoring on the active interval $\mathcal{F}$ rather than on the full waveform. The MRSTFT reward is $R_{\mathrm{MRSTFT}}=-\log(1+\ell_{\mathrm{MRSTFT}})$, where $\ell_{\mathrm{MRSTFT}}$ is the multi-resolution short-time Fourier transform loss. Lower spectral loss therefore gives a higher reward.

The reward components have different scales. We therefore standardize each component over the local batch $\mathcal{B}$ and combine them as
\begin{equation}
r_{g,i}
=
\sum_m
w_m
\frac{
R_{g,i}^{(m)}-\mu_{\mathcal{B}}^{(m)}
}{
s_{\mathcal{B}}^{(m)}
},
\label{eq:reward-scalar}
\end{equation}
where the sum is over the active components for that run, and $\mathcal{B}$ contains all retained rollouts from all mixtures in the current update. For component $m$, $\mu_{\mathcal{B}}^{(m)}$, $s_{\mathcal{B}}^{(m)}$, and $w_m$ are its batch mean, batch standard deviation, and weight, respectively. We use $w_m=1$ for every component. Eq.~\ref{eq:advantage} then normalizes $r_{g,i}$ within group $g$.

\textbf{Sample-quality filtering.} Before computing the reward, the filter applies the target-level, active-duration, clipping, channel-alignment, and channel-balance thresholds in Section~\ref{sec:exp}. The checks remove targets that are low-level, inactive, clipped, channel-misaligned, or channel-imbalanced. On these targets, a waveform reward can be poorly conditioned.

\section{Experimental Setup}
\label{sec:exp}

\textbf{Data.}
We use $50$k internal mixtures, each $20$ seconds long. We hold out $50$ mixtures as the test set and exclude them from training. We use a pretrained audio VAE, $\epsilon$ar-VAE~\cite{earvae}, and freeze its parameters.

\textbf{Model.}
The pretrained Flow Matching model is a 1B-parameter continuous Transformer with $24$ layers, hidden dimension $1536$, and $24$ attention heads. FlowGRPO updates only this model.

\textbf{Training.}
Each run uses $5$k optimization steps with one inner update per step and a batch size of $4$ mixtures. Training uses two NVIDIA A100 GPUs with $80$ GB of memory each. We use AdamW~\cite{adamw} with a learning rate of $2\times 10^{-6}$ and weight decay of $10^{-3}$. Sampling follows Section~\ref{sec:flowgrpo}. Each rollout group contains $G=8$ trajectories conditioned on the same mixture and initialized from the same noise. We use $N=10$, $a=0.7$, and an SDE window of length $K=2$. The start index $n_0$ is sampled uniformly from $\{1,\ldots,5\}$, so the window stays on earlier steps. This follows the finding that early-step exploration is more informative~\cite{tempflowgrpo}. We set $\epsilon=0.2$ and $w_m=1$.

The sample-quality thresholds below are fixed independently of the pre-training coefficients in Table~\ref{tab:reward_gap}. When filtering is enabled, targets must have an RMS of at least $-25\,\mathrm{dBFS}$ and at least $3\,\mathrm{s}$ of activity above $-40\,\mathrm{dBFS}$. Their clipping ratio, channel-onset lag, and channel-RMS ratio must not exceed $0.01$, $30\,\mathrm{ms}$, and $3$, respectively.

\textbf{Metrics.}
We report ViSQOL~\cite{visqol}, SI-SDR~\cite{sisdr}, signal-to-distortion ratio (SDR)~\cite{sdr_bss_eval}, PESQ~\cite{pesq}, and the MRSTFT loss $\ell_{\mathrm{MRSTFT}}$~\cite{mrstft_loss}. Higher values are better for the first four metrics, whereas lower MRSTFT is better. ViSQOL always denotes the original evaluation metric, including when $\mathrm{ViSQOL}_{\mathrm{act}}$ is used as a training reward.

\section{Results}

Table~\ref{tab:reward_ablation} reports vocal-separation quality. We make three controlled comparisons: FlowGRPO post-training against the pretrained separator, the revised ViSQOL against the original definition under a fixed filter, and quality-based filtering against no filtering under a fixed reward. Higher values are better for all metrics except the MRSTFT loss. The ViSQOL column reports the original evaluation metric, even when training uses $\mathrm{ViSQOL}_{\mathrm{act}}$.

\textbf{Ablation results.}
Rows 1 and 2 show that MRSTFT-only FlowGRPO improves every reported metric over the pretrained separator, with the lowest MRSTFT loss among all five rows. Rows 4 and 5 use the same filter and differ only in the ViSQOL definition. Using $\mathrm{ViSQOL}_{\mathrm{act}}$ instead of the original ViSQOL improves all five evaluation metrics. Similarly, Rows 3 and 5 use the same revised reward and differ only in sample-quality filtering. Compared with Row 3, filtering improves all five metrics in Row 5, whose PESQ also exceeds the pretrained value in Row 1. Overall, Row 5 achieves the highest ViSQOL, SI-SDR, SDR, and PESQ, while Row 2 retains the lowest MRSTFT loss.

\textbf{Training dynamics.}
Fig.~\ref{fig:reward_curves} shows the reward trajectories during training. All three runs exhibit increasing rewards, indicating that FlowGRPO successfully optimizes the respective training objectives. The increases in Rows 4 and 5 are substantially larger than that in Row 3, suggesting that sample-quality filtering facilitates GRPO optimization. Moreover, Row 5 exhibits a slightly larger increase than Row 4, suggesting that the revised ViSQOL further benefits the optimization. Notably, reward values in Fig.~\ref{fig:reward_curves} are not directly comparable across settings for two reasons: data quality is associated with reward values, and enabling the filter changes the quality distribution of training samples; additionally, the reward definitions differ across settings. Therefore, final separation performance should be assessed using the evaluation metrics in Table~\ref{tab:reward_ablation}.

\section{Conclusion}

In this work, we study reward design for FlowGRPO post-training in vocal separation. Before policy optimization, we analyze the association between within-group reward gaps and target data quality factors. This analysis motivates a normalized active-interval ViSQOL and a sample-quality filter based on waveform checks. Experiments demonstrate that the proposed reward combination substantially improves separation quality and perceptual quality. Importantly, reward design can be guided by rollouts from a fixed model before policy optimization. This work highlights the value of pre-optimization reward analysis, which future work will extend to more Flow Matching tasks and reward functions.


\begin{thebibliography}{10}

\bibitem{diffusion_source_separation}
Robin Scheibler, Youna Ji, Soo-Whan Chung, Jaeuk Byun, Soyeon Choe, and
  Min-Seok Choi.
\newblock Diffusion-based generative speech source separation.
\newblock In {\em Proc. IEEE International Conference on Acoustics, Speech and
  Signal Processing (ICASSP)}, pages 1--5, 2023.

\bibitem{diffusion_speech_enhancement}
Julius Richter, Simon Welker, Jean-Marie Lemercier, Bunlong Lay, and Timo
  Gerkmann.
\newblock Speech enhancement and dereverberation with diffusion-based
  generative models.
\newblock {\em IEEE/ACM Transactions on Audio, Speech, and Language
  Processing}, 31:2351--2364, 2023.

\bibitem{gen_target_speaker_extraction}
Theodor Nguyen, Guangzhi Sun, Xianrui Zheng, Chao Zhang, and Philip~C.
  Woodland.
\newblock Conditional diffusion model for target speaker extraction.
\newblock {\em arXiv preprint arXiv:2310.04791}, 2023.

\bibitem{fm_audio_1}
Shivam Mehta, Ruibo Tu, Jonas Beskow, {\'E}va Sz{\'e}kely, and Gustav~Eje
  Henter.
\newblock {Matcha-TTS}: A fast {TTS} architecture with conditional flow
  matching.
\newblock In {\em Proc. IEEE International Conference on Acoustics, Speech and
  Signal Processing (ICASSP)}, pages 11341--11345, 2024.

\bibitem{fm_audio_2}
Yi~Yuan, Xubo Liu, Haohe Liu, Mark~D. Plumbley, and Wenwu Wang.
\newblock {FlowSep}: Language-queried sound separation with rectified flow
  matching.
\newblock In {\em Proc. IEEE International Conference on Acoustics, Speech and
  Signal Processing (ICASSP)}, pages 1--5, 2025.

\bibitem{lipman2023flow}
Yaron Lipman, Ricky T.~Q. Chen, Heli Ben-Hamu, Maximilian Nickel, and Matthew
  Le.
\newblock Flow matching for generative modeling.
\newblock In {\em International Conference on Learning Representations}, 2023.

\bibitem{flowgrpo}
Jie Liu, Gongye Liu, Jiajun Liang, Yangguang Li, Jiaheng Liu, Xintao Wang,
  Pengfei Wan, Di~Zhang, and Wanli Ouyang.
\newblock Flow-{GRPO}: Training flow matching models via online {RL}.
\newblock In {\em Advances in Neural Information Processing Systems},
  volume~38, pages 45650--45685, 2025.

\bibitem{wer_training}
Rohit Prabhavalkar, Tara~N. Sainath, Yonghui Wu, Patrick Nguyen, Zhifeng Chen,
  Chung-Cheng Chiu, and Anjuli Kannan.
\newblock Minimum word error rate training for attention-based
  sequence-to-sequence models.
\newblock In {\em Proc. IEEE International Conference on Acoustics, Speech and
  Signal Processing (ICASSP)}, pages 4839--4843, 2018.

\bibitem{valle}
Chengyi Wang, Sanyuan Chen, Yu~Wu, Ziqiang Zhang, Long Zhou, Shujie Liu, Zhuo
  Chen, Yanqing Liu, Huaming Wang, Jinyu Li, Lei He, Sheng Zhao, and Furu Wei.
\newblock Neural codec language models are zero-shot text to speech
  synthesizers.
\newblock {\em arXiv preprint arXiv:2301.02111}, 2023.

\bibitem{metricgan}
Szu-Wei Fu, Chien-Feng Liao, Yu~Tsao, and Shou-De Lin.
\newblock {MetricGAN}: Generative adversarial networks based black-box metric
  scores optimization for speech enhancement.
\newblock In {\em International Conference on Machine Learning}, pages
  2031--2041, 2019.

\bibitem{sdr_bss_eval}
E.~Vincent, R.~Gribonval, and C.~F{\'e}votte.
\newblock Performance measurement in blind audio source separation.
\newblock {\em IEEE Transactions on Audio, Speech, and Language Processing},
  14(4):1462--1469, 2006.

\bibitem{vae}
Diederik~P. Kingma and Max Welling.
\newblock Auto-encoding variational {Bayes}.
\newblock In {\em International Conference on Learning Representations}, 2014.

\bibitem{koizumi2017}
Yuma Koizumi, Kenta Niwa, Yusuke Hioka, Kazunori Kobayashi, and Yoichi Haneda.
\newblock {DNN}-based source enhancement self-optimized by reinforcement
  learning using sound quality measurements.
\newblock In {\em Proc. IEEE International Conference on Acoustics, Speech and
  Signal Processing (ICASSP)}, 2017.

\bibitem{koizumi2018}
Yuma Koizumi, Kenta Niwa, Yusuke Hioka, Kazunori Kobayashi, and Yoichi Haneda.
\newblock {DNN}-based source enhancement to increase objective sound quality
  assessment score.
\newblock {\em IEEE/ACM Transactions on Audio, Speech, and Language
  Processing}, 26(10):1780--1792, 2018.

\bibitem{f5rtts}
Xiaohui Sun, Ruitong Xiao, Jianye Mo, Bowen Wu, Qun Yu, and Baoxun Wang.
\newblock {F5R-TTS}: Improving flow-matching based text-to-speech with group
  relative policy optimization.
\newblock {\em arXiv preprint arXiv:2504.02407}, 2025.

\bibitem{grpo}
Zhihong Shao, Peiyi Wang, Qihao Zhu, Runxin Xu, Junxiao Song, Xiao Bi, Haowei
  Zhang, Mingchuan Zhang, Y.~K. Li, Y.~Wu, and Daya Guo.
\newblock {DeepSeekMath}: Pushing the limits of mathematical reasoning in open
  language models.
\newblock {\em arXiv preprint arXiv:2402.03300}, 2024.

\bibitem{visqol}
Michael Chinen, Felicia S.~C. Lim, Jan Skoglund, Nikita Gureev, Feargus
  O'Gorman, and Andrew Hines.
\newblock {ViSQOL} v3: An open source production ready objective speech and
  audio metric.
\newblock In {\em International Conference on Quality of Multimedia Experience
  (QoMEX)}, pages 1--6, 2020.

\bibitem{sisdr}
Jonathan {Le Roux}, Scott Wisdom, Hakan Erdogan, and John~R. Hershey.
\newblock {SDR} -- half-baked or well done?
\newblock In {\em Proc. IEEE International Conference on Acoustics, Speech and
  Signal Processing (ICASSP)}, pages 626--630, 2019.

\bibitem{pesq}
A.~W. Rix, J.~G. Beerends, M.~P. Hollier, and A.~P. Hekstra.
\newblock Perceptual evaluation of speech quality ({PESQ}) -- a new method for
  speech quality assessment of telephone networks and codecs.
\newblock In {\em Proc. IEEE International Conference on Acoustics, Speech and
  Signal Processing (ICASSP)}, pages 749--752, 2001.

\bibitem{diffwave}
Zhifeng Kong, Wei Ping, Jiaji Huang, Kexin Zhao, and Bryan Catanzaro.
\newblock {DiffWave}: A versatile diffusion model for audio synthesis.
\newblock In {\em International Conference on Learning Representations}, 2021.

\bibitem{wavegrad}
Nanxin Chen, Yu~Zhang, Heiga Zen, Ron~J. Weiss, Mohammad Norouzi, and William
  Chan.
\newblock {WaveGrad}: Estimating gradients for waveform generation.
\newblock In {\em International Conference on Learning Representations}, 2021.

\bibitem{gradtts}
Vadim Popov, Ivan Vovk, Vladimir Gogoryan, Tasnima Sadekova, and Mikhail
  Kudinov.
\newblock {Grad-TTS}: A diffusion probabilistic model for text-to-speech.
\newblock In {\em International Conference on Machine Learning}, pages
  8599--8608, 2021.

\bibitem{zuo2026decoupled}
Lishi Zuo, Youzhi Tu, Lu Yi, Zezhong Jin, Chongxin Gan, Man-Wai Mak, and
  KongAik Lee.
\newblock Decoupled Latent Flow Matching for Few-Step Joint
  Vocal-Accompaniment Separation.
\newblock {\em arXiv preprint arXiv:2608.30913}, 2026.

\bibitem{demucs}
Alexandre D{\'e}fossez, Nicolas Usunier, L{\'e}on Bottou, and Francis Bach.
\newblock Music source separation in the waveform domain.
\newblock {\em arXiv preprint arXiv:1911.13254}, 2019.

\bibitem{bsrnn}
Yi~Luo and Jianwei Yu.
\newblock Music source separation with band-split {RNN}.
\newblock {\em IEEE/ACM Transactions on Audio, Speech, and Language
  Processing}, 31:1893--1901, 2023.

\bibitem{bsroformer}
Wei-Tsung Lu, Ju-Chiang Wang, Qiuqiang Kong, and Yun-Ning Hung.
\newblock Music source separation with band-split {RoPE} transformer.
\newblock In {\em Proc. IEEE International Conference on Acoustics, Speech and
  Signal Processing (ICASSP)}, pages 481--485, 2024.

\bibitem{diff_singing_vocal}
Gen{\'i}s Plaja-Roglans, Yun-Ning Hung, Xavier Serra, and Igor Pereira.
\newblock Generating separated singing vocals using a diffusion model
  conditioned on music mixtures.
\newblock In {\em Proc. IEEE Workshop on Applications of Signal Processing to
  Audio and Acoustics (WASPAA)}, pages 1--5, 2025.

\bibitem{prismaudio}
Huadai Liu, Kaicheng Luo, Wen Wang, Qian Chen, Peiwen Sun, Rongjie Huang,
  Xiangang Li, Jieping Ye, and Wei Xue.
\newblock {PrismAudio}: Decomposed chain-of-thoughts and multi-dimensional
  rewards for video-to-audio generation.
\newblock In {\em International Conference on Learning Representations}, 2026.

\bibitem{tempflowgrpo}
Xiaoxuan He, Siming Fu, Yuke Zhao, Wanli Li, Jian Yang, Dacheng Yin, Fengyun
  Rao, and Bo~Zhang.
\newblock {TempFlow-GRPO}: When timing matters for {GRPO} in flow models.
\newblock {\em arXiv preprint arXiv:2508.04324}, 2025.

\bibitem{flowttsgrpo}
Haoxu Wang, Biao Tian, Weiqin Li, Xiang Lv, Han Zhao, and Xiangang Li.
\newblock {FlowTTS-GRPO}: Online reinforcement learning with multi-objective
  reward optimization for flow-matching based text-to-speech.
\newblock In {\em Proc. Interspeech}, 2026.

\bibitem{reward_hacking_1}
Joar Skalse, Nikolaus H.~R. Howe, Dmitrii Krasheninnikov, and David Krueger.
\newblock Defining and characterizing reward hacking.
\newblock In {\em Advances in Neural Information Processing Systems},
  volume~35, pages 9460--9471, 2022.

\bibitem{reward_hacking_2}
Alexander Pan, Kush Bhatia, and Jacob Steinhardt.
\newblock The effects of reward misspecification: Mapping and mitigating
  misaligned models.
\newblock In {\em International Conference on Learning Representations}, 2022.

\bibitem{earvae}
Kangdi Wang, Zhiyue Wu, Dinghao Zhou, Rui Lin, Junyu Dai, and Tao Jiang.
\newblock Back to ear: Perceptually driven high fidelity music reconstruction.
\newblock {\em arXiv preprint arXiv:2509.14912}, 2025.

\bibitem{adamw}
Ilya Loshchilov and Frank Hutter.
\newblock Decoupled weight decay regularization.
\newblock In {\em International Conference on Learning Representations}, 2019.

\bibitem{mrstft_loss}
Ryuichi Yamamoto, Eunwoo Song, and Jae-Min Kim.
\newblock Parallel {WaveGAN}: A fast waveform generation model based on
  generative adversarial networks with multi-resolution spectrogram.
\newblock In {\em Proc. IEEE International Conference on Acoustics, Speech and
  Signal Processing (ICASSP)}, pages 6199--6203, 2020.

\end{thebibliography}
\end{document}